\documentclass[a4paper]{article}

\usepackage{icrc2013}
\usepackage{amssymb,amsmath}
\usepackage{lscape}
\usepackage{longtable}
\usepackage{ctable}
\usepackage{multirow}
\usepackage{siunitx}
\usepackage[english]{babel}

\title{Time-dependent modelling of pulsar wind nebulae}

\shorttitle{Time-dependent modelling of PWNe}

\authors{
M.J. Vorster$^{1}$,
O. Tibolla$^{2}$,
S.E.S. Ferreira$^{1}$,
and S. Kaufmann$^{3}$
}

\afiliations{
$^1$ Centre for Space Research, North-West University, Potchefstroom Campus, 2520 Potchefstroom, South Africa \\
$^2$ Institut f\"ur Theoretische Physik und Astrophysik, Universit\"at W\"urzburg, Campus Hubland Nord, Emil-Fischer-Str. 31, D-97074 W\"urzburg, Germany \\
$^3$ Landessternwarte, Universit\"at Heidelberg, K\"onigstuhl 12, D-69117 Heidelberg, Germany \\
}

\email{12792322@puk.ac.za}

\abstract{
A spatially-independent model that can be used to calculate the temporal evolution of the electron/positron spectrum in a spherically expanding pulsar wind nebula is presented. The model is applied to the young nebula G21.5-0.9, as well as to the unidentified TeV sources HESS J1427-608 and HESS J1507-622.  The parameters derived from the model strengthens the idea that the unidentified sources can be identified as evolved pulsar wind nebulae.
}

\keywords{ISM: individual objects (G21.5-0.9, HESS J1427-608, HESS J1507-622) - ISM: supernova remnants - radiation mechanisms: non-thermal}

\begin{document}
\maketitle

\section{Introduction}


It is well known that the X-ray synchrotron emission observed from pulsar wind nebulae (PWN) is produced by a young population of electrons, as these particles have a relatively short lifetime.  By contrast, the electrons producing very high energy (VHE) gamma-ray emission through the inverse Compton (IC) scattering of background photons have a much longer lifetime, implying that the observed TeV emission from PWNe is produced by particles that have accumulated over the lifetime of the pulsar \cite{Dejager2009}.  For a PWN with an average magnetic field of $B=5\,\mu\,\text{G}$, the lifetime of an electron emitting $1\,\text{keV}$ X-rays is $\sim 3\,\text{kyr}$, whereas the corresponding lifetime of an electron producing $1\,\text{TeV}$ gamma-rays is $\sim 19\,\text{kyr}$ \cite{Dejager2009}.

Although PWNe are often identified based on the characteristics of their radio and X-ray synchrotron spectra (e.g., \cite{Dejager2009,Weiler1980}), it has been proposed by \cite{Dejager2008a} that the magnetic field in an evolved PWN could evolve below the value of the interstellar medium (ISM), resulting in these sources being undetectable at synchrotron frequencies.  However, due to the longer lifetimes of the VHE gamma-ray producing electrons, these ancient PWNe may still be visible at TeV energies.  As PWNe form a significant fraction of detected TeV sources, the ancient PWN scenario could offer an explanation for a number of unidentified TeV sources \cite{H_Aharonian2008} that lack a clear synchrotron counterpart. 

In this paper we present time-dependent modelling of the candidate PWNe HESS J1427-608 \cite{H_Aharonian2008} and HESS J1507-622 \cite{H_Aharonian2011}.  The aim is to not only investigate the ancient PWN hypothesis, but also to determine whether a clear argument can be made for identifying HESS J1427-608 and HESS J1507-622 as PWNe.  Before presenting the modelling results, the spatially-independent model used to calculate the temporal evolution of the electron spectrum is introduced.  In order to test the model, it is first applied to the young PWN G21.5-0.9.  

For a more extended discussion of the results, \cite{Vorster2013} should be consulted.  Using the present model, predictions for additional unidentified TeV sources have also been made by \cite{Tibolla2013}.

\section{The Model}\label{sec:model}

The temporal evolution of the electron spectrum in a PWN can be calculated using the equation (e.g., \cite{Tanaka2010})
\begin{equation}\label{eq:dn_dt}
\frac{\partial N_e(E_e,t)}{\partial t} = Q(E_e,t) + \frac{\partial}{\partial E}\left[\dot{E}(E_e,t)N_e(E_e,t)\right], 
\end{equation}
where $E_e$ represents the electron energy and $N_e(E_e,t)$ the number of electrons per energy interval.  The number of electrons injected into the PWN at the termination shock, per time and energy interval, is given by $Q(E_e,t)$, while the second term on the right-hand side of (\ref{eq:dn_dt}) describes continuous energy losses (or gains) suffered by the particles. The value $\dot{E}(E_e,t)$ represents the total energy loss rate as a result of the various processes.  

Emulating \cite{Venter2006}, a broken power-law spectrum is used to model the emission from the sources studied in this paper
\begin{equation}\label{eq:power-law}
Q(E_e,t) = \begin{cases}
Q_{\rm{R}}\left(E_{\rm{b}}/E_e\right), & \text{if } E_{\min} \le E_e \le E_{\rm{b}}\\
Q_{\rm{X}}\left(E_{\rm{b}}/E_e\right)^2, & \text{if } E_{\rm{b}} < E_e \le E_{\max} 
\end{cases},
\end{equation}
where $Q_{\rm{R}}$ and $Q_{\rm{X}}$ are normalisation constants, $E_{\min}$ and $E_{\max}$ the minimum and maximum electron energy respectively, and $E_{\rm{b}}$ the energy where the spectrum transitions between the two components.  With the results of \cite{Dejager2008c} taken into account, it is not an a priori requirement that the two components should have the same intensity at $E_{\rm{b}}$.

The normalisation constants are determined by the prescription that the total energy in a given source component should be some fraction $\eta_i$ ($i=\mbox{ R,X}$) of the pulsar's spin-down luminosity $L(t)$ \cite{Venter2006}
\begin{equation}\label{eq:Q_calc}
\int Q_i\left(E_{\rm{b}}/E_e\right)^{p_i} E_e dE_e = \eta_i L(t),
\end{equation}
with 
\begin{equation}\label{eq:L_t}
L(t) = \frac{L_0}{\left(1+t/\tau\right)^2}.
\end{equation}
In the expression above $L_0$ is the initial luminosity and $\tau$ the characteristic spin-down time scale of the pulsar, while it is assumed that the pulsar is a pure dipole radiator with a braking index of $3$.

The expressions given in \cite{Blumenthal1970} are used to calculate the non-thermal energy losses that result from synchrotron radiation and IC scattering in the Thomson regime.  During the modelling it was found that the energy density of the IC target photons becomes larger than the energy density of the magnetic field, and it thus becomes necessary to include a Klein-Nishina correction to the non-thermal energy loss rate.  The approximations derived by \cite{Moderski2005} is used for this purpose.  Following \cite{Tanaka2010}, adiabatic cooling (heating) is described in the present model by
\begin{equation}\label{eq:E_dot_ad2}
\dot{E}_{\rm{ad}}(E_e,t) = \frac{v_{\rm{pwn}}(t)}{R_{\rm{pwn}}(t)}E_e, 
\end{equation}
where $v_{\rm{pwn}}(t)$ and $R_{\rm{pwn}}(t)$ are respectively the expansion velocity and radius of the PWN.  An important parameter in the model is the evolution of $R_{\rm{pwn}}(t)$.  Not only is this required for the adiabatic energy loss rate (\ref{eq:E_dot_ad2}), but also for the evolution of the average magnetic field and escape loss rate (as will be discussed below).  

Simulations predict that the evolution of a PWN can be divided into three phases: in the initial evolutionary phase the nebula expands at a rate $R_{\rm{pwn}}(t) \propto t^{r_1}$ (e.g., \cite{Reynolds1984,Gelfand2009}).  When the pressure in the supernova shell remnant (SNR) becomes small enough, the reverse shock of the shell will start to propagate towards the centre of the remnant \cite{McKee1974}, reaching the outer boundary of the PWN after a time $t_{\rm{rs}}$, and the next evolutionary phase begins.  The reverse shock compresses the PWN over a time scale of a few thousand years (e.g., \cite{Vanderswaluw2001, Bucciantini2003}), leading to a stronger magnetic field and enhanced synchrotron losses \cite{Reynolds1984}.  After the initial compression, the PWN will enter a second expansion phase starting at $t=t_{\rm{se}}$.  In contrast to the first expansion phase, $R_{\rm{pwn}}(t)$ does not evolve smoothly, but has an oscillatory nature \cite{Vanderswaluw2001, Bucciantini2003}.  As a first approximation, all three evolutionary phases of $R_{\rm{pwn}}(t)$ are modelled using a power-law 
\begin{equation}\label{eq:R_t}
R_{\rm{pwn}}(t) = \left\{ \begin{array}{ll}
R_0(t/t_0)^{r_1} & \textrm{if $t<t_{\rm{rs}}$}\\
R_0(t_{\rm{rs}}/t_0)^{r_1}(t/t_{\rm{rs}})^{r_2} & \textrm{if $t_{\rm{rs}}\le t<t_{\rm{se}}$}\\
R_0(t_{\rm{rs}}/t_0)^{r_1}(t_{\rm{se}}/t_{\rm{rs}})^{r_2}(t/t_{\rm{se}})^{r_3} & \textrm{if $t\ge t_{\rm{se}}$}
\end{array}  \right.
\end{equation}
where $R_0=0.01$ pc is the initial radius at $t_0=10$ yr \cite{Gelfand2009}.  The values $r_1$, $r_2$, and $r_3$ are not linearly independent as the size of the PWN predicted by the model must be equal to the observed size.  Note that the distance to the source $d$ influences the values of $r_1$, $r_2$ and $r_3$, as a larger value of $d$ implies a larger source, and hence a faster expansion.  The expansion velocity of the PWN follows from the usual definition $v_{\rm{pwn}}(t)= dR_{\rm{pwn}}(t)/dt$.

An expression for $t_{\rm{rs}}$ has been derived by a number of authors, including \cite{Reynolds1984, Ferreira2008}.  It is estimated that an SNR with an ejecta mass of $M_{\rm{ej}}=5 M_{\odot}$, and a kinetic energy of $E_{\rm{ej}}=10^{51}$ erg, expanding into an ISM with a density of $n_{\rm{ism}}= 1 \mbox{ cm}^{-3}$, will have a reverse shock time scale of $t_{\rm{rs}} \approx 6000$ yr.  Using a smaller value for the ISM density ($n_{\rm{ism}}= 0.1 \mbox{ cm}^{-3}$) increases the reverse shock time scale to $t_{\rm{rs}} \approx 11000$ yr.   

The evolution of the average magnetic field in the nebula $B_{\rm{pwn}}(t)$ is calculated using the conservation of magnetic flux \cite{Tanaka2010}
\begin{equation}\label{eq:B_evolve}
\int_0^{t} \eta_B L(t)dt = V_{\rm{pwn}}(t)\frac{B^2_{\rm{pwn}}(t)}{8\pi},
\end{equation} 
where $\eta_B$ is the fraction of the pulsar's spin-down luminosity converted into magnetic energy, and $V_{\rm{pwn}}(t)$ the volume of the PWN.  

The ratio of electromagnetic to particle energy in the nebula $\sigma$ is defined in the model as 
\begin{equation}
\sigma = \frac{\eta_B}{\eta_{\rm{R}}+\eta_{\rm{X}}},
\end{equation}
with the constraint $\sigma \lesssim 1$ imposed \cite{Dejager2009}.  An additional constraint is supplied by $\eta_B+\eta_{\rm{R}}+\eta_{\rm{X}} \lesssim 1$.  This sum is not set strictly equal to unity to allow for the fact that a fraction $\eta_{\rm{rad}}$ of the spin-down luminosity is radiated away in the form of pulsed emission.  For the modeling, $\eta_{\rm{rad}} \lesssim 1\%$ is used.  Note that this small value used for $\eta_{\rm{rad}}$ effectively implies that $\sigma \simeq \eta_B$.

Particles can also escape from the PWN as a result of diffusion.  The escape time scale $\tau_{\rm{esc}}$ is given by
\begin{equation}\label{eq:tau_esc}
\tau_{\rm{esc}} = \frac{R^2_{\rm{pwn}}(t)}{6\kappa(t)},
\end{equation}
where $\kappa(t)$ is the diffusion coefficient.  Diffusion in a PWN results from particles interacting with irregularities in the magnetic field, and the scaling $\kappa(t) \propto 1/B_{\rm{pwn}}(t)$ is used in the model.  The diffusion coefficient is also proportional to energy, i.e., $\kappa\equiv \kappa (E_e/1\mbox{ TeV})$.  

The energy loss equation (\ref{eq:dn_dt}) is not solved directly, but rather using the approximation method described in \cite{Vorster2013}.  The non-thermal emission is calculated using the appropriate expressions given in \cite{Blumenthal1970}.  This implies that Klein-Nishina effects are taken into account when calculating the IC spectrum.

\section{Results}


\begin{table}[ht]
\begin{center}
	\begin{tabular}{lccc}
	\hline\midrule
 	Parameter & G21.5 & J1427 & J1507 \\
  \midrule
  $L_0$ ($10^{38}\,\text{erg}\,\text{s}^{-1}$) & $0.54*$ & $5.5$ & $1.2$ \\
  $\tau$ (kyr) & $3^*$ & $3$ & $0.5$ \\
  $t_{\rm{age}}$ (kyr) & $0.87^*$ & $10$ & $24$ \\
  $B_{\rm{age}}$ ($\mu\,\text{G}$) & $230$ & $0.4$ & $1.7$  \\
  $\eta_{\rm{R}}$ & $0.68$ & $0.81$ & $0.8$ \\
  $\eta_{\rm{X}}$ & $0.13$ & $0.18$ & $0.17$ \\
  $\eta_{\rm{R}}/\eta_{\rm{X}}$ & $5.4^*$ & $4.5^*$ & $4.7^*$ \\
  $\sigma$ ($10^{-3}$) & $180$ & $0.01$ & $30$  \\
  $E_{\min}$ ($10^{-3}\,\text{TeV}$) & $0.3$ & $100$ & $1$ \\
  $E_{\min}$ ($10^2\,\text{TeV}$) & $2.7$ & $3$ & $2$ \\
  $E_{\rm{b}}$ (TeV) & $0.1$ & $0.18$ & $0.5$ \\  
  $\kappa_{\rm{age}}$ ($10^{25} E_{\rm{TeV}}\,\text{cm}^2\,\text{s}^{-1}$) & $2.2$ & $7$ & $15$\\
  $\kappa/\kappa_{\rm{Bohm}}$ & $390^*$ & $2.3^*$ & $20^*$\\
  $d$ (kpc) & $4.8^*$ & $11^*$ & $6^*$ \\
  \midrule
\end{tabular}
\caption{Values derived with the model for the various free parameters.  Values marked with an * represent parameters that were kept fixed, or parameters that follow from the derived model parameters.  The ratio $\kappa/\kappa_{\rm{Bohm}}$ expresses the derived diffusion coefficient in terms of the Bohm diffusion coefficient.}
\label{tab:parameters}
\end{center}
\end{table}

\subsection{G21.5-0.9}

 \begin{figure}[!t]
  \centering
  \includegraphics[width=0.5\textwidth]{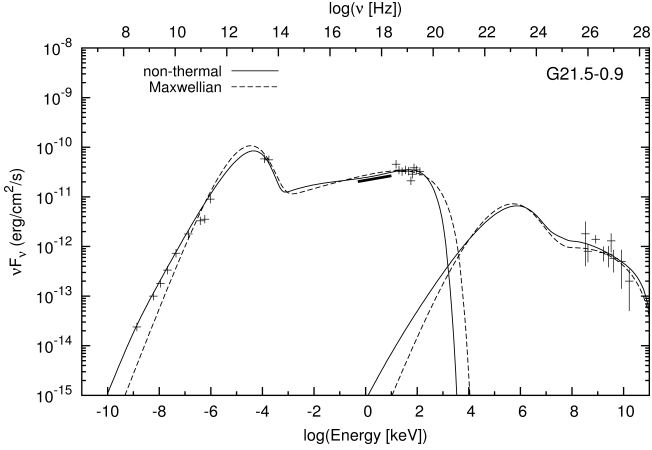}
  \caption{Model prediction for G21.5-0.9.  Radio data are taken from \cite{Goss1970, Becker1975, Morsi1987, Salter1989a, Salter1989b, Bock2001, Bandiera2001, Bietenholz2008, Bietenholz2011}, infra-red data from \cite{Gallant1999}, X-ray data from \cite{Slane2000, DeRosa2009} and the INTEGRAL Science Data Centre (http://www.isdc.unige.ch/heavens webapp/integral), and TeV data from \cite{Dejager2008b}.  The data were also modelled using the Maxwellian spectrum calculated by \cite{Spitkovsky2008}.}
  \label{fig:G21.5}
 \end{figure}

With a spin-down luminosity of $L=3.3 \times 10^{37}\,\text{erg}\,\text{s}^{-1}$ \cite{Camilo2006}, the pulsar in the supernova remnant G21.5-0.9 is one of the most energetic pulsars in the Galaxy.  The PWN is located at a distance of $4.8\,\text{kpc}$ \cite{Tian2008}, with an estimated age of $870\,\text{yr}$ \cite{Bietenholz2008}.  For the modeling of G21.5-0.9, the value $\tau=3\,\text{kyr}$ is used \cite{Dejager2009b}.  The PWN is too young to have interacted with the reverse shock, and must therefore still be in the first expansion phase.      

Figure \ref{fig:G21.5} shows the model prediction for the non-thermal radiation spectra, with the derived parameters listed in Table \ref{tab:parameters}.  From the model a present-day magnetic field of $B_{\rm{age}}=230\,\mu\,\text{G}$ is derived.  This is comparable to the value of $B_{\rm{age}}=300\,\mu\,\text{G}$ derived for the $\sim 1\,\text{kyr}$ old Crab nebula.  To obtain the model prediction, the diffusion coefficient should not be larger than $\kappa= 2.2 \times 10^{25} E_{\rm{TeV}}\,\text{cm}^2\,\text{s}^{-1}$, or in terms of the Bohm diffusion coefficient, $\kappa=390\kappa_{\rm{Bohm}}$.  Furthermore, a ratio of $\eta_{\rm{R}}/\eta_{\rm{X}}=5.4$ is derived for the conversion efficiencies.  From the model prediction a value of $\sigma=0.18$ is derived for the ratio of magnetic to particle energy, larger than the value of $\sigma \sim 0.003$ calculated for the Crab Nebula \cite{Kennel1984b}.  


\vspace{0.5cm}

\subsection{HESS J1427-608}

\begin{figure}[!t]
  \centering
  \includegraphics[width=0.5\textwidth]{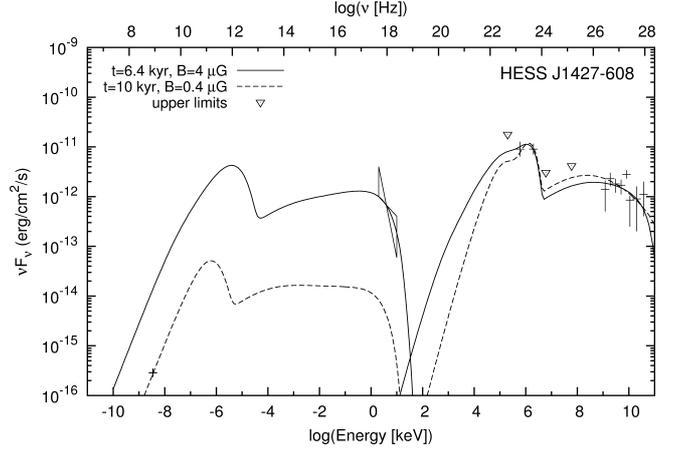}
  \caption{Model prediction for the unidentified TeV source HESS J1427-608.  The radio data is taken from \cite{Murphy2007}, the X-ray data from \cite{Fujinaga2012}, the \emph{Fermi} data from \cite{Nolan2012}, and the TeV data from \cite{H_Aharonian2008}.  The parameters listed in Table \ref{tab:parameters} are for the $B_{\rm{age}}=0.4\,\mu\,\text{G}$ scenario.}
  \label{fig:1427}
 \end{figure}

One of the unidentified TeV sources discovered in a H.E.S.S. galactic plane survey is HESS J1427-608 \cite{H_Aharonian2008}.  From the model prediction an initial luminosity of $L_0=5.5\times 10^{38}\,\text{erg}\,\text{s}^{-1}$ is derived, leading to a present-day luminosity of $L=2.9 \times 10^{37}\,\text{erg}\,\text{s}^{-1}$.  The ratios $\eta_{\rm{R}}/\eta_{\rm{X}}=4.5$ and $\sigma =10^{-5}$ are derived, along with a present-day magnetic field of $B_{\rm{age}}=0.4\,\mu\,\text{G}$.  Lastly, a diffusion coefficient of $\kappa= 5.6 \times 10^{25} E_{\rm{TeV}}\,\text{cm}^2\,\text{s}^{-1}$ is predicted by the model, or equivalently, $\kappa=2.3\kappa_{\rm{Bohm}}$.  While these parameters may lead to an acceptable agreement between the model and radio data, Figure \ref{fig:1427} shows that this scenario significantly under-predicts the \emph{Suzaku} spectrum \cite{Fujinaga2012}.  It is possible to predict the X-ray observations using a larger magnetic field of $B_{\rm{age}}=4\,\mu\,\text{G}$.  However, this scenario predicts a bright radio synchrotron nebula that has thus far not been observed.  The fact that the model cannot simultaneously predict both the radio and X-ray observations indicates that one of the two synchrotron sources is not a plausible counterpart to HESS J1427-608.  Note that the parameters listed in Table \ref{tab:parameters} are specifically for the $B_{\rm{age}}=0.4\,\mu\,\text{G}$ scenario      
    
\vspace{0.5cm}

\subsection{HESS J1507-622}

Also discovered in a H.E.S.S. Galactic plane survey is the bright VHE source HESS J1507-622 \cite{H_Aharonian2011}.  This source is unique in the sense that it lies $\sim 3.5^{\circ}$ from the Galactic plane, whereas all other unidentified TeV sources lie within $\pm 1^{\circ}$ from the Galactic equator.  Most Galactic VHE sources are connected to young stellar populations (located in the disk), and one would therefore not expect a bright VHE source at the observed position.  Furthermore, the absence of a bright X-ray counterpart is surprising as the comparably low hydrogen column density at $\sim 3.5^{\circ}$ leads to a considerably lower absorption of X-rays, as well as reduced background emission \cite{H_Aharonian2011}.  

Assuming that HESS J1507-622 is still in the initial expansion phase, a break energy of $E_{\rm{b}}=0.5\,\text{TeV}$ is derived.  This is an order of magnitude larger than the value derived for other PWNe \cite{Zhang2008, Tanaka2011}.  Although it is not excluded that such a large break energy is the result of shock acceleration, an alternative scenario is favoured in the present paper where HESS J1507-622 has been compressed by the reverse shock of the SNR.  As $\dot{E_e}/E_e$ is constant in (\ref{eq:E_dot_ad2}), the effect of adiabatic losses is to shift the electron spectrum to lower energies without affecting the spectral shape.  During the compression phase, the exact opposite will occur.  The particles will be subjected to adiabatic heating, causing the electron spectrum to shift to higher energies, leading to an increase in the value of $E_{\rm{b}}$.  

\begin{figure}[!t]
  \centering
  \includegraphics[width=0.5\textwidth]{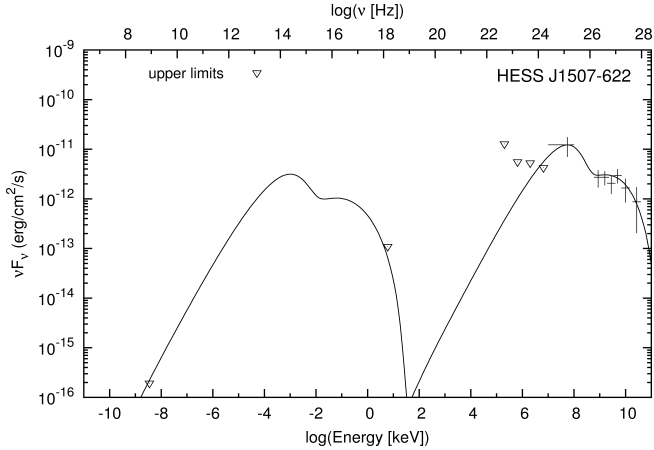}
  \caption{Model prediction for the unidentified source HESS J1507-622.  The radio upper limit is taken from \cite{Green1999}, and the GeV data from \cite{Nolan2012}.  The X-ray upper limit and TeV data are taken from \cite{H_Aharonian2011}. }
  \label{fig:1507}
 \end{figure}

For the compression scenario, the value of $t_{\rm{rs}} = 20\,\text{kyr}$ is used, while the compression phase is chosen to last for $4\,\text{kyr}$.  It is assumed that the nebula has not yet entered the second expansion phase, so that the current age of the PWN is $t_{\rm{age}}=24\,\text{kyr}$.  Given the offset from the Galactic plane, it seems reasonable to assume that the ISM is homogeneous.  This will lead to a symmetric reverse shock and a preservation of the spherical nature of the PWN, even after the interaction with the reverse shock.  The model prediction resulting from the compression scenario is shown in Figure \ref{fig:1507}, with the derived parameters listed in Table \ref{tab:parameters}.  With the compression taken into account, the break energy is reduced to $E_{\rm{b}}=0.5\,\text{TeV}$.  Note that this is the break energy of the source spectrum at the pulsar wind termination shock, while the particle spectrum in the PWN has a break at $E_e=5\,\text{TeV}$.  

Other parameters derived include a relatively large initial luminosity ($L_0=1.2\times 10^{38}\,\text{erg}\,\text{s}^{-1}$) and short spin-down time scale ($\tau=0.5\,\text{kyr}$).  A present-day magnetic field of $B_{\rm{age}}=1.7\,\mu\,\text{G}$ is derived from the model, larger than the value of $B_{\rm{age}}=0.5\,\mu\,\text{G}$ estimated by \cite{H_Aharonian2011}.  Additionally, the ratios $\eta_{\rm{R}}/\eta_{\rm{X}}=4.7$ and $\sigma=0.03$ are derived, as well as $\kappa=1.5 \times 10^{26} E_{\rm{TeV}}\,\text{cm}^2\,\text{s}^{-1}$, or $\kappa=20\kappa_{\rm{Bohm}}$.  

\section{Summary and conclusions}

In this paper a time-dependent particle evolution model for PWNe is presented. This model is applied to the nebula G21.5-0.9, as well as to the unidentified TeV sources HESS J1427-608 and HESS J1507-622.  For the three sources parameters are derived that are reasonable within a PWN framework, thereby strengthening the argument that HESS J1427-608 and HESS J1507-622 can be identified as evolved PWNe.  

Motivated by observations, a broken power-law is used as the source spectrum for the electrons injected into the PWN at the termination shock.  In contrast to previous PWN models of a similar nature (e.g., \cite{Tanaka2010,Zhang2008}), the source spectrum in the present model has a discontinuity in intensity at the transition between the low and high-energy components.  The choice of a discontinuous source spectrum leads to a better model prediction of the data at all wavelengths, in contrast to a continuous one.  A similar conclusion has also been drawn by \cite{Dejager2008c} from their modelling of Vela X.  As a discontinuous spectrum is also required for the young ($t_{\rm{age}}\sim 1\,\text{kyr}$) nebula G21.5-0.9, the discrepancy between the two components cannot be an artefact of PWN evolution.  A characteristic of the discontinuous spectrum is that a particle conversion efficiency must be specified for both the low ($\eta_{\rm{R}}$) and high-energy ($\eta_{\rm{X}}$) components, with a ratio of $\eta_{\rm{R}}/\eta_{\rm{X}}\sim 4.5-5.4$ derived for the three sources.  

The research presented in this paper is discussed in more detail by \cite{Vorster2013}, with additional modelling results of unidentified TeV sources presented by \cite{Tibolla2013}.


\end{document}